\documentclass[preprint,times,12pt]{elsarticle}
\usepackage{amsmath,graphicx,url}
\newcommand{\bX}{\mathbf{X}}
\newcommand{\bx}{\mathbf{x}}
\newcommand{\by}{\mathbf{y}}
\newcommand{\bz}{\mathbf{z}}
\begin{document}
\title{Macroeconomic Phase Transitions Detected from the Dow Jones
Industrial Average Time Series}
\author[mas]{WONG Jian Cheng}
\author[mas]{LIAN Heng}
\author[pap]{CHEONG Siew Ann\corref{cor1}}
\ead{cheongsa@ntu.edu.sg}

\cortext[cor1]{Corresponding author}

\address[mas]{Division of Mathematical Sciences,
School of Physical and Mathematical Sciences,
Nanyang Technological University,
21 Nanyang Link, Singapore 637371,
Republic of Singapore}

\address[pap]{Division of Physics and Applied Physics,
School of Physical and Mathematical Sciences,
Nanyang Technological University,
21 Nanyang Link, Singapore 637371,
Republic of Singapore}

\begin{abstract}
In this paper, we perform statistical segmentation and clustering
analysis of the Dow Jones Industrial Average time series between
January 1997 and August 2008.  Modeling the index movements and
log-index movements as stationary Gaussian processes, we find a total
of 116 and 119 statistically stationary segments respectively.  These
can then be grouped into between five to seven clusters, each
representing a different macroeconomic phase.  The macroeconomic
phases are distinguished primarily by their volatilities.  We find the
US economy, as measured by the DJI, spends most of its time in a
low-volatility phase and a high-volatility phase.  The former can be
roughly associated with economic expansion, while the latter contains
the economic contraction phase in the standard economic cycle.  Both
phases are interrupted by a moderate-volatility market, but
extremely-high-volatility market crashes are found mostly within the
high-volatility phase.  From the temporal distribution of various
phases, we see a high-volatility phase from mid-1998 to mid-2003, and
another starting mid-2007 (the current global financial crisis).
Transitions from the low-volatility phase to the high-volatility phase
are preceded by a series of precursor shocks, whereas the transition
from the high-volatility phase to the low-volatility phase is preceded
by a series of inverted shocks.  The time scale for both types of
transitions is about a year.  We also identify the July 1997 Asian
Financial Crisis to be the trigger for the mid-1998 transition, and an
unnamed May 2006 market event related to corrections in the Chinese
markets to be the trigger for the mid-2007 transition.

\end{abstract}


\maketitle

\section{Introduction}

Most people remember the most recent economic recession as short
(lasting only eight months from March 2001 to November 2001
\cite{NBER}) and mild (affecting mostly high-tech companies).  Against
this backdrop, there have been many sensationalist claims that the
current global financial crisis is the deepest (broad spectrum of
economic sectors affected) and longest (peak in December 2007
\cite{NBER}, and a potential trough in March 2009).  According to
other sources (see, for example, Ref.~\cite{Wiki}), however, the
Subprime Crisis surfaced around July 2007 with a slew of bad news from
subprime lenders, and the Dow Jones Industrial Average (DJI) dipping
roughly 1,000 points going from July 2007 to August 2007.  Since then,
billions of dollars have been sunk into relief and stimulus packages,
and governments around the world are planning further aid totalling in
excess of a trillion US dollars.  There are hardly any positive
results to show for the effort thus far, and the reasons can best be
summed up as ``too little, too late''.  In medicine, early
intervention is generally more effective and less costly compared to a
late cure.  The same is probably true for economies and financial
markets.  Clearly, even if we are not sure what kind of intervention
measures will be effective, acting early is still more desirable to
acting later.  To accomplish this, it is important to be able to
unambiguously detect the onset of a financial crisis, so that we can
at the same time avoid over-reacting when the market has merely caught
a `cold'.

Since econometric data such as the gross national product (GNP) are
released quarterly, and are adjusted monthly, they are not useful for
timely detection.  We thus look to higher-frequency financial time
series for this sleuth work.  Given that each and every financial
crisis may have their own unique and esoteric characters, we need a
financial time series that is sufficiently representative of the broad
spectrum of industries to be able to detect the starting point of
these crises.  Indices such as the Dow Jones Industrial Average (DJI),
Dow Jones Composite Average (DJA), and the Standards \& Poors 500
(INX) are most suitable for this purpose.  Clearly, detecting the
onset of a financial crisis is a change point problem
\cite{Carlstein1994ChangePointProblems,
Chen2000ParametricStatisticalChangePointAnalysis}.  In their seminal
works, Goldfeld \emph{et al}, Hamilton and Kim \emph{et al} fitted a
Markov-switching model to local trends in the US GNP time series to
detect transitions between a macroeconomic \emph{phase} (or
\emph{regime}) with high growth rate and a macroeconomic phase with
low growth rate \cite{Goldfeld1973JEconometrics1p3,
Hamilton1989Econometrica57p357, Kim1999RevEconoStat81p608}.  Unlike
econometric time series, which evolve fairly slowly with time, it is
well known that financial time series exhibit dynamics on multiple
time scales.  To avoid potential complications arising from such
multiscale dynamics, we analyze statistical fluctuations in the index
time series, instead of looking merely at the local trend, as is done
for deciding the duration of an economic recession.  For the different
macroeconomic phases the economy and financial market can be found in,
these statistical flucutations should also be qualitatively different. 

In this paper, we describe in Section \ref{sect:models} a model-based
approach to statistically segmenting the DJI time series, which is
assumed to consist of a large number of statistically stationary
\emph{segments}.  Within different segments, the index movements (or
log index movements) are assumed to follow stationary Gaussian
processes with different means and variances.  We then discover these
segments using a recursive segmentation scheme based on the relative
entropy between them.  Following this, we determine the small number
of macroeconomic phases represented in the time series by performing
agglomerative hierarchical clustering on the segments.  In Section
\ref{sect:results}, we report findings from our statistical
segmentation and clustering analyses.  Segments obtained using the two
models are in good agreement with each other, and also with the dates
of major market events, suggesting that the segment boundaries
discovered are robust and meaningful.  Depending on the model, and the
level of granularity we choose, we find between six to seven
macroeconomic phases after clustering the segments.  These six to
seven macroeconomic phases are distinguished primarily by their
variances, which represent market volatilities.  While the clusters
appear to be less robust compared to the segments, their temporal
distributions do tell a fairly consistent story: the US market, as
measured by the DJI, is found predominantly in a low-volatility phase
and a high-volatility phase, corresponding roughly to economic
expansion and economic contraction respectively.  Both phases are
interrupted by a moderate-volatility market correction phase, while
the high-volatility phase is also interrupted by an
extremely-high-volatility market crash phase.  More interestingly, our
results suggest that the mid-1998 transition into the high-volatility
phase (which lasted five years) was triggered by the 1997 Asian
Financial Crisis, whereas the mid-2007 transition into the
high-volatility phase (the global financial crisis we find ourselves
in right now) was triggered by a 2006 correction in the Chinese
markets.  As we have guessed, the world is very tightly coupled
economically, perhaps even more so than we would like to admit.  We
then conclude in Section \ref{sect:conclusions}, and describe further
work we are currently undertaking.

\section{Data, Models and Methods}
\label{sect:models}

\subsection{Data and Models}

While it is not as comprehensive as the S\&P 500, the Dow Jones
Industrial Average (a price-weighted index consisting of 30 of the
largest and most widely held public companies in US) is nonetheless a
very important index measuring the performance of the US market.
Tic-by-tic data for this index between 1 January 1997 and 31 August
2008 was downloaded from the Taqtic database \cite{Taqtic}, and
processed to give a half-hourly time series $\bX = (X_1, X_2, \dots,
X_N)$, where $X_t$ is the index value at the $t$th half-hour, and $N$
is the total number of trading half-hours between 1 January 1997 to 31
August 2008.  The half-hourly frequency was chosen so that there is
sufficient statistics to identify segments as short as a single day.
From the index time series $\bX$, we obtain the index movement time
series $\bx = (x_1, \dots, x_{N-1})$, where $x_t = X_t - X_{t-1}$, as
well as the log-index movement time series $\by = (y_1, \dots,
y_{N-1})$, where $y_t = \log X_t - \log X_{t-1}$.  We assume that
$\bx$ and $\by$ consist of $M$ and $M'$ statistically stationary
segments respectively, where the numbers of segments $M$ and $M'$, and
where the segments are, are unknown and must be determined through a
segmentation procedure.  

To do this segmentation, we assume the movements $x_t$ within
statistically stationary segment $m$ are drawn from a \emph{Gaussian}
(\emph{normal}) distribution with mean $\mu_m$ and variance
$\sigma_m^2$.  Similarly, the movements $y_t$ within statistically
stationary segment $m'$ are assumed to be drawn from a Gaussian
distribution with mean $\mu'_{m'}$ and variance ${\sigma'_{m'}}^2$.  The
log-normal index movement model is popular in the finance literature,
where traders are assumed to be influenced mainly by percentage
changes rather than absolute changes, because of their constant mental
reference to a risk-free interest rate.  In this study, we also
consider the normal index movement model, in case traders in the real
world also pay attention to actual changes in the index.  In both
models, the movements from one half-hour to the next are uncorrelated,
in contrast to real-world financial time series, which are known to
exhibit correlations on multiple time scales.  For the purpose of
finding statistically robust change points in the time series, we
believe that the details of the models used will not be important, and
the difference between an uncorrelated model versus a correlated model
will merely be a difference between statistical significance and
signal-to-noise ratio.

\subsection{Methods}

Time series segmentation schemes can be very broadly classified into
those based on pattern recognition, and those based on
information-theoretic measures.  In pattern-based segmentation
schemes, features within the time series are abstracted into symbols,
as is frequently done in the technical analysis of stock markets
\cite{Murphy1999}.  Segmentation decisions are then based on the
relative abundance of symbols, or their context trees
\cite{Chung2002ICDM2002p83, Jiang2007ProcWiCom2007p5609,
Xie2007IntJInfoSysSci3p479,
Zhang2007LecNoteCompSci14798p520}.  Information-theoretic segmentation
methods are popular in image segmentation
\cite{BarrancoLopez1995ElectronicLett31p867}, biological sequence
segmentation \cite{Braun1998StatSci13p142}, and also in medical time
series analysis \cite{BernaolaGalvan2001PhysRevLett87a168105}, but not
widely used for financial time series segmentation
\cite{Oliver1998LecNoteCompSci1394p222, Lemire2006CS0605103}.

To determine the location of the $M$ segments, we employ the recursive
segmentation scheme introduced by Bernaola-Galv\'an \emph{et al}
\cite{BernaolaGalvan1996PhysicalReviewE53p5181,
RomanRoldan1998PhysicalReviewLetters80p1344} for biological sequence
segmentation.  In this scheme, we first identify a
\emph{cursor position} $t$ in the sequence $\bz = (z_1, z_2, \dots, z_N)$
with length $N$, and compute the Jensen-Shannon divergence
\begin{equation}
\Delta_t = \log\frac{P_2(t)}{P_1},
\end{equation}
which measures the statistical divergence between the left
subsequence $\bz_L = (z_1, z_2, \dots, z_t)$ and the right subsequence
$\bz_R = (z_{t+1}, z_{t+2}, \dots, z_N)$.  Here,
\begin{equation}
P_1 = \prod_{i=1}^N \frac{1}{\sqrt{2\pi \sigma^2}}
\exp\left[-\frac{(z_i - \mu)^2}{2\sigma^2}\right]
\end{equation}
is the likelihood for observing the sequence $\bz = (z_1, z_2, \dots,
z_N)$, assuming that the entire sequence is generated by a single 
Gaussian process with mean $\mu$ and variance $\sigma^2$, and
\begin{equation}
P_2(t) = \prod_{i=1}^t \frac{1}{\sqrt{2\pi \sigma_L^2}}
\exp\left[-\frac{(z_i - \mu_L)^2}{2\sigma_L^2}\right]
\prod_{j=t+1}^N \frac{1}{\sqrt{2\pi \sigma_R^2}}
\exp\left[-\frac{(z_j - \mu_R)^2}{2\sigma_R^2}\right]
\end{equation}
is the likelihood for observing the sequence $\bz$, assuming that the
left subsequence $\bz_L = (z_1, z_2, \dots, z_t)$ is generated by a
Gaussian process with mean $\mu_L$ and variance $\sigma_L^2$, and the
right subsequence $\bz_R = (z_{t+1}, z_{t+2}, \dots, z_N)$ is
generated by a Gaussian process with mean $\mu_R$ and variance
$\sigma_R^2$.

Since the parameters $\mu$, $\mu_L$, $\mu_R$, $\sigma^2$,
$\sigma_L^2$, and $\sigma_R^2$ are not given, we can replace them with
their maximum-likelihood estimates $\hat{\mu}$, $\hat{\mu}_L$,
$\hat{\mu}_R$, $\hat{\sigma}^2$, $\hat{\sigma}_L^2$, and
$\hat{\sigma}_R^2$.  These estimates maximizes $P_1$ and $P_2(t)$
relative to the data, and the Jensen-Shannon divergence, which
simplifies to
\begin{equation}
\Delta_t = N\log\hat{\sigma} - n_L \log\hat{\sigma}_L - n_R
\log\hat{\sigma}_R + \frac{1}{2} \geq 0,
\end{equation}
tells us how much better the best two-segment model fits the observed
data over the best one-segment model.  If we now vary $t$, and
identify $t = t^*$ for which $\Delta_{t^*} = \Delta^* = \max_t
\Delta_t$, this would tell us the best place to segment the given
sequence $\bz = (z_1, z_2, \dots, z_N)$.  The Jensen-Shannon
divergence maximum $\Delta^*$ gives us an indication of how
significant the segment boundary at $t^*$ is statistically.

We then repeat this one-into-two segmentation procedure to recursively
cut the given sequence up into shorter and shorter segments.  As this
recursive segmentation progresses, the divergence maxima for the new
cuts will generally become smaller and smaller.  At some point, new
cuts will no longer be statistically significant, and the segmentation
process must be terminated.  There are several ways to do this:
through hypothesis testing
\cite{BernaolaGalvan1996PhysicalReviewE53p5181,
RomanRoldan1998PhysicalReviewLetters80p1344}, through model selection,
\cite{Li2001PhysicalReviewLetters86p5815, Li2001ProcRECOMB01p204}, or
through examination of the intrinsic statistical fluctuations within
the sequence to be segmented \cite{CheongIRJSS}.  In this work, we
adopted a semi-automated approach to terminate the recursive
segmentation.  First, we recursively segment the time series until the
divergence maxima of the new cuts fall below a given threshold,
selected by inspection to be $\Delta_0 = 10$.  We then screen these
segments manually, by visually inspecting the Jensen-Shannon
divergence spectrum $\Delta_t$, to decide whether very short segments
should be eliminated, and very long segments should be further
segmented.

At each stage of the recursive segmentation, we also perform
segmentation optimization, to overcome the \emph{context sensitivity
problem} identified in Ref.~\cite{CheongCSP}.  For this, we use the
algorithm described in Ref.~\cite{CheongIRJSS}, where we start with
$M$ segment boundaries $\{t_1, \dots, t_M\}$ obtained after new cuts
have been introduced by the recursive segmentation.  To optimize the
position of the $m$th segment boundary, we compute the Jensen-Shannon
divergence spectrum $\Delta_t$ within the \emph{supersegment}
$(x_{t_{m-1}+1}, \dots, x_{t_{m+1}})$ bounded by the segments
boundaries $t_{m-1}$ and $t_{m+1}$, and replace $t_m$ by $t_m^*$,
where the supersegment Jensen-Shannon divergence is maximized.  This
is done for all $M$ segment boundaries, and iterated until all segment
boundaries converge to their optimal positions.  We then continue the
recursive segmentation with this optimized set of segments,
introducing new cuts, optimize the new segment boundaries along with
the old segment boundaries, until the segmentation is terminated.

Finally, after we are satisfied that the final segmentation is
optimal, and the segment boundaries are all statistically significant,
we perform agglomerative hierarchical clustering on the segments to
determine the number of macroeconomic phases represented in the time
series.  This is done with the complete link algorithm
\cite{Jain1999}, using the Jensen-Shannon divergences between segments
as their statistical distances.  Clustering of different periods
within a financial time series has been previously investigated
\cite{vanWijk1999ProcInfoVisualp4, Krawiecki2002PhysRevLett89e158701,
Fu2004ProcIntConfDataMiningp5}, but we believe we are the first to
incorporate a rigorous segmentation analysis into such a study.

\section{Results and Discussions}
\label{sect:results}

\subsection{Statistical Segmentation}

From the DJI time series between January 1997 and August 2008, we
found a total of 116 segments using the normal index movement model,
and a total of 119 segments for the log-normal index movement model.
Most of the optimized segment boundaries found are either mid-days or
end-of-days, in agreement with the start-of-day and end-of-day buzz,
and mid-day lull observed in practically all financial markets
\cite{Admati1988RevFinStudies1p3, Gourieroux1999JFinMarkets2p193}.  We
say that a segment boundary is \emph{common} between the two sets if
its positions in the two models differ by at most one day.  A total of
85 common segment boundaries are found, out of which 37 are at the
same exact half-hour.  This tells us that most of the segment
boundaries discovered are extremely robust.  As shown in Figure
\ref{fig:DJIsegments}, these robust segment boundaries agree very well
with the dates of important market events.  In Table
\ref{table:discord}, we also show the intervals where the segmentations
from the the two models disagree.  These intervals are bound by very
robust segment boundaries, and most of these intervals correspond to
highly volatile periods in the DJI time series.  Within these
intervals, disagreement between the two models is primarily in the
form of different number of segment boundaries.  We surmise that the
statistical fluctuations within these intervals are highly
nonstationary, and thus not well described by a collection of
stationary models.  Even so, we find many common segment boundaries
within these intervals.

\begin{table}[htbp]
\centering\footnotesize
\caption{Intervals within the January 1997 to August 2008 period where
segmentations of the normal index movement model and log-normal
index movement model disagree.}
\label{table:discord}
\vskip .5\baselineskip
\begin{tabular}{ccccc}
\hline
\raisebox{-6pt}[0pt][0pt]{start date} & 
\raisebox{-6pt}[0pt][0pt]{end date} & 
\multicolumn{2}{c}{number of segments} &
common \\
\cline{3-4}
& & \parbox[t]{2.5cm}{\centering normal index movement model} & 
\parbox[t]{2.5cm}{\centering log-normal index movement model\strut} & 
boundaries \\
\hline
Nov 3, 1997 & Mar 31, 1998 & 4 & 5 & 0 \\
Aug 26, 1998 & Oct 20, 1998 & 3 & 2 & 0 \\
Jan 13, 1999 & Nov 5, 1999 & 3 & 7 & 0 \\
Mar 9, 2001 & Jun 3, 2002 & 18 & 10 & 6 \\
Oct 16, 2002 & Aug 6, 2003 & 9 & 6 & 2 \\
Mar 10, 2004 & Oct 18, 2005 & 3 & 8 & 1 \\
Jul 28, 2006 & Aug 15, 2006 & 1 & 2 & 0 \\
Sep 5, 2006 & Dec 27, 2006 & 4 & 1 & 0 \\
Jul 25, 2007 & Mar 10, 2008 & 7 & 14 & 4 \\
\hline
\end{tabular}
\end{table}

\subsection{Statistical Clustering}

\begin{figure}[htbp]
\centering\footnotesize
(normal index movement model)
\vskip .5\baselineskip
\includegraphics[scale=0.29]{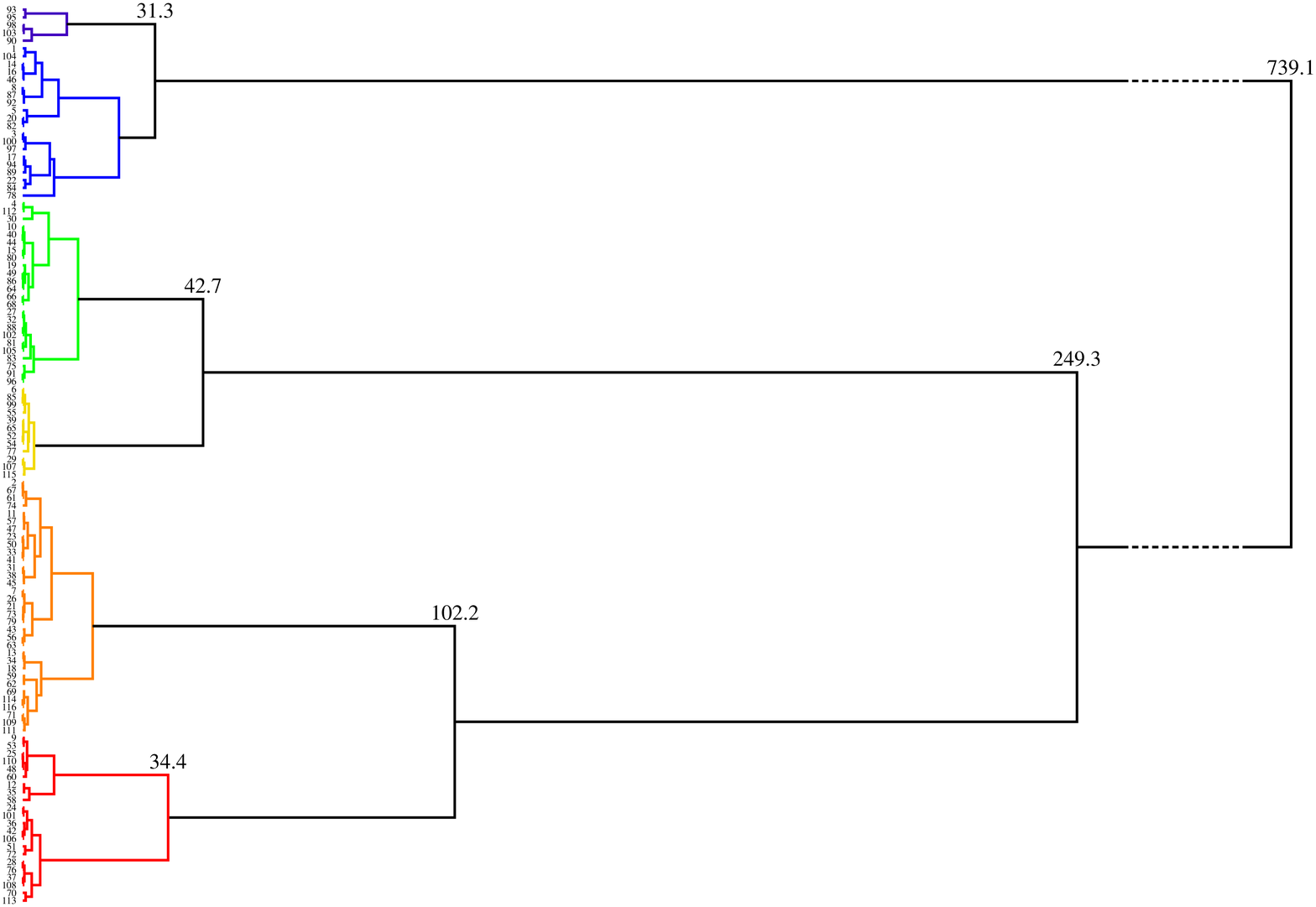}
\vskip\baselineskip
(log-normal index movement model)
\vskip .5\baselineskip
\includegraphics[scale=0.29]{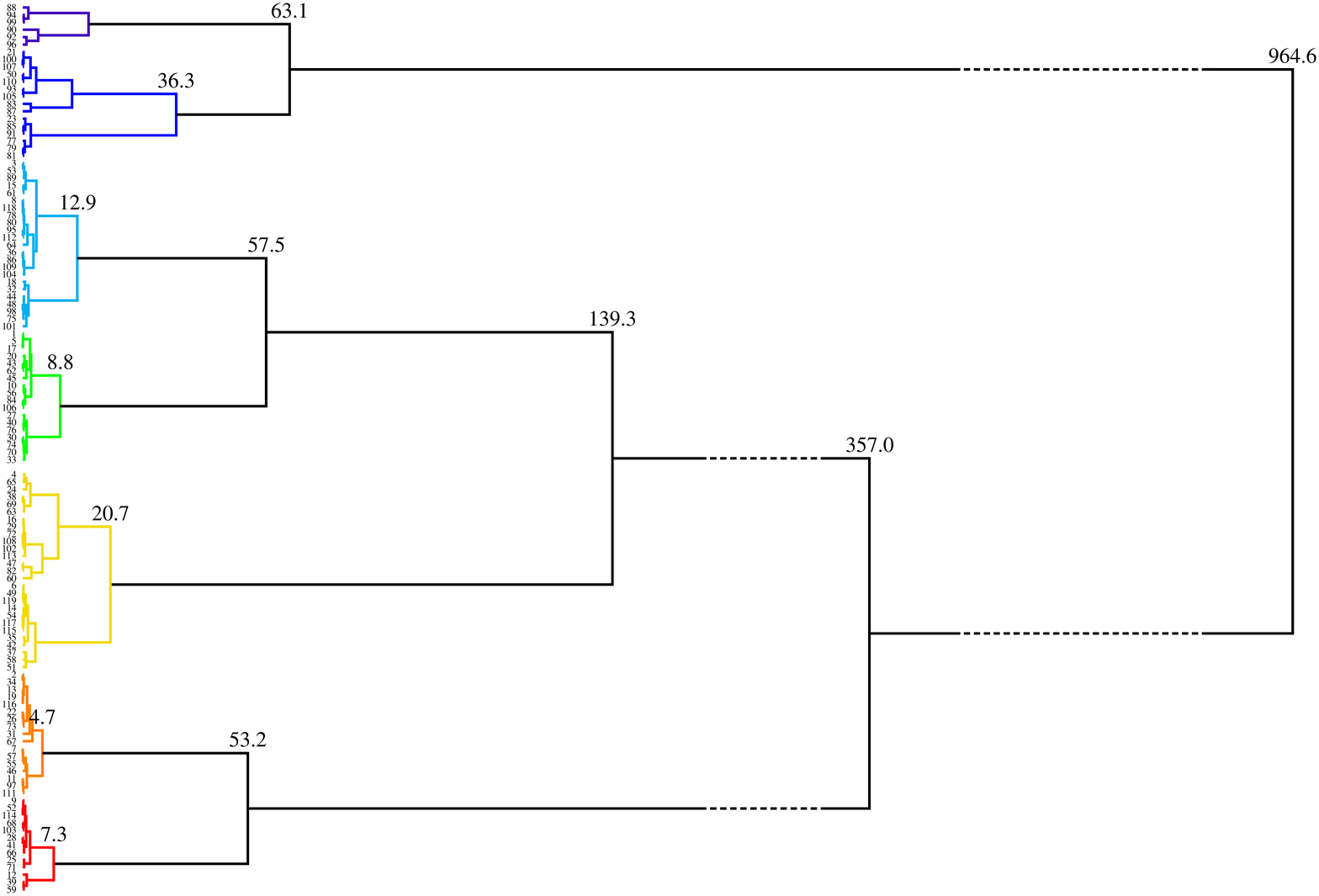}
\caption{The complete-link hierarchical clustering trees of the
segments obtained using the normal index movement model (top) and the
log-normal index movement model (bottom). The differentiated clusters
are coloured according to their market volatilities: low (deep blue
and blue), moderate (cyan and green), high (yellow and orange), and
extremely high (red).  Also shown at the major branches are the 
Jensen-Shannon divergence values at which subclusters are merged.}
\label{figure:trees}
\end{figure}

In their classic studies \cite{Goldfeld1973JEconometrics1p3,
Hamilton1989Econometrica57p357}, Goldfeld \emph{et al} and Hamilton
assumed only two macroeconomic phases for the US GNP.  More recently,
Sims and Zha assumed four phases in their analysis of the history of
US monetary policy \cite{Sim2006AmEconRev96p54}.  In general,
economists believe in the existence of only a small number of
macroeconomic phases.  On the large scale, the textbook economic cycle
consists of recurrent switches between an economic expansion phase and
an economic contraction phase.  On a smaller scale, economists also
acknowledge the existence of a market correction phase and a market
crash phase.  Based on our clustering analysis of the segments, we
find indeed a small number of clusters, as shown in Figure
\ref{figure:trees}.  For the normal index movement model, we find
between five to seven clusters of segments, depending on the level of
granularity we choose.  Similarly, the hierarchical clustering tree of
the log-index movement model suggests seven clusters of segments.  For
both models, the coarsest description that is reasonable and
informative is in terms of three clusters of segments.

\begin{figure}[htbp]
\centering
\includegraphics[scale=0.5, clip=true]{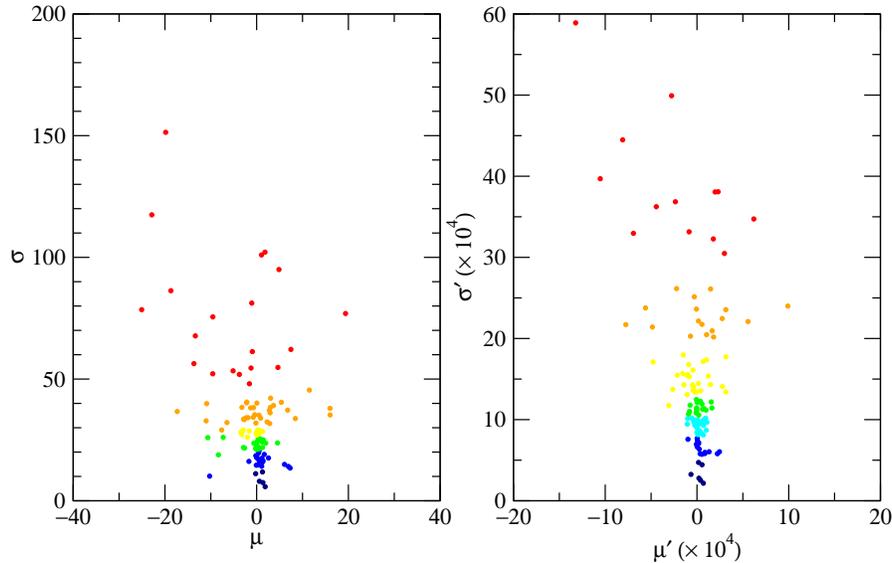}
\caption{Means and standard deviations of the segments obtained using
the normal index movement model (left) and the log-normal index
movement model (right).  As we can see, the clusters are
differentiated primarily by their standard deviations.}
\label{fig:musigmascatter}
\end{figure}

\begin{figure}[htbp]
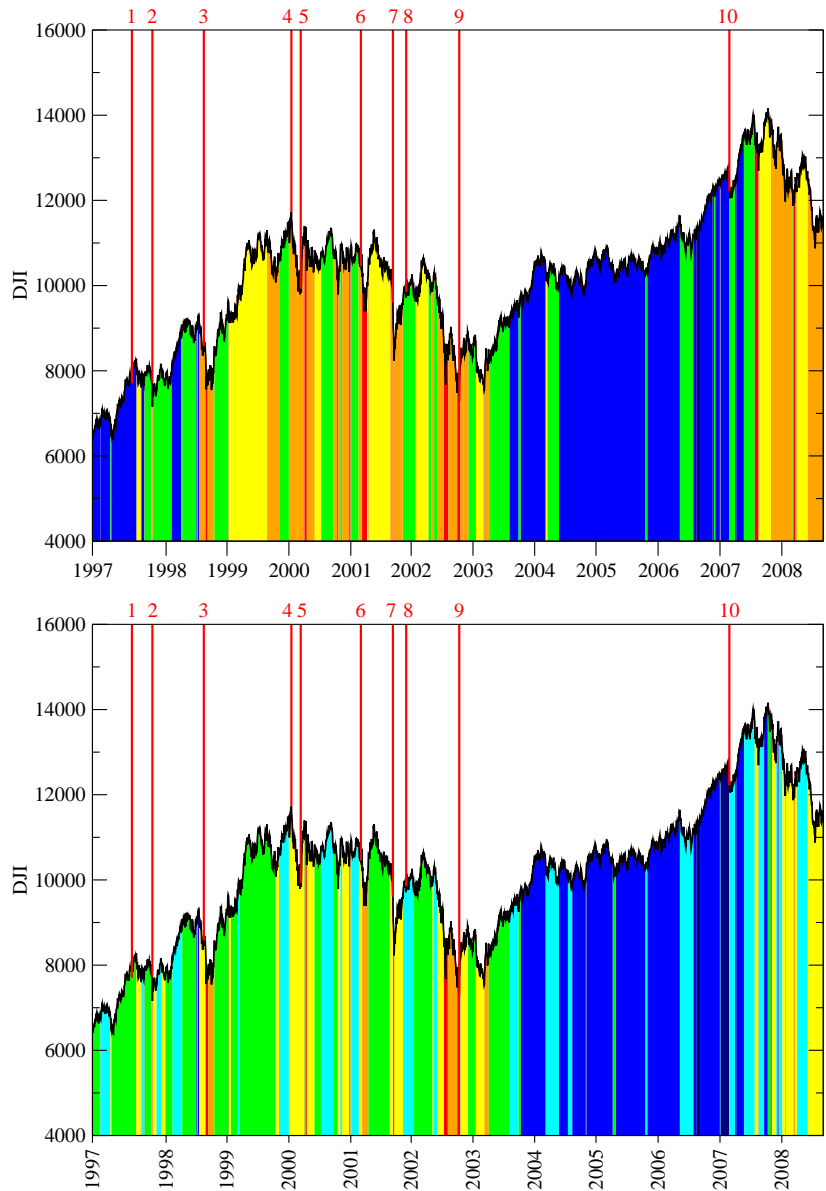

\centering
\includegraphics[scale=0.45, clip=true]{DJIJan1997Aug2008}
\vskip .5\baselineskip
\includegraphics[scale=0.45, clip=true]{logDJIJan1997Aug2008}
\caption{Temporal distributions of the clustered segments for the
normal index movement model (top), and the log-normal index movement
model (bottom).  The red solid lines indicate the dates of important
market events: (1) July 1997 Asian Financial Crisis; (2) October 1997
Mini Crash; (3) August 1998 Russian Financial Crisis; (4) DJI 2000
High; (5) NASDAQ Crash; (6) start of 2001 recession; (7) Sep 11
Attack; (8) end of 2001 recession; (9) DJI 2002 Low; (10) February
2007 Chinese Correction.}
\label{fig:DJIsegments}
\end{figure}

When we plot a scatter diagram of the segment means and standard
deviations, as shown in Figure \ref{fig:musigmascatter}, we see that
the clusters are distinguished primarily through their standard
deviations, i.e. their market volatilities.  Adopting a heat-map-like
colour scheme for the clusters, we colour the low-volatility clusters
deep blue and blue, the moderate-volatility clusters cyan and green,
the high-volatility clusters yellow and orange, and the
extremely-high-volatility clusters red.  Using this colour scheme, we
plot the temporal distributions of clustered segments for the two
models as Figure \ref{fig:DJIsegments}.  The two temporal
distributions agree qualitatively on the existence of a low-volatility
phase between mid-2003 to end-2006, and a high-volatility phase within
2008.  However, we find the log-normal index movement model
exaggerates small statistical divergences, at the same time playing
down large statistical divergences.  As such, there is higher temporal
contrast at low market volatilities, and lower temporal contrast at
high market volatilities.  In comparison, the normal index movement
model, with its uniform contrast between market volatilities, tells us
a much more interesting story: over the period January 1997 to August
2008, the US market, as measured by the DJI, is found predominantly in
the low-volatility (deep blue and blue) and high-volatility (yellow
and orange) phases.  By visual inspection of the DJI time series, we
see that the low-volatility phase has a natural interpretation as the
economic expansion phase, but while the high-volatility phase contains
the economic contraction phase, its duration is significantly longer.
From this point on, we will limit our discussions to the normal index
movement model.

As we can see from Figure \ref{fig:DJIsegments}, both the
low-volatility phase and the high-volatility phase are interrupted by
a moderate-volatility market correction phase (green).  In the normal index
movement model, segments within this phase have very consistent
standard deviations of about 20 index points.  The length distribution of these
market correction segments, however, is bimodal, with one group
lasting between 100--200 half-hours (1--2 weeks), and another group
lasting between 700--900 half-hours (1.5--2 months).  In general, we
find more short correction segments within the low-volatility phase,
and more long correction segments within the high-volatility phase.
The high-volatility phase is also interrupted
frequently by an extremely-high-volatility market crash phase, which
sports a broad range of standard deviations from 50 to 150 index
points.  Crash segment lengths were also found to fall into three
groups: between 10--40 half-hours (1--3 days), around 100 half-hours
(1 week), and between 200--300 half-hours (2--3 weeks).

\subsection{Temporal Distribution of Clustered Segments}

Most importantly, the temporal distribution of the clustered segments
between January 1997 and August 2008 indicates the US market made a
transition from the low-volatility phase to the high-volatility phase
in mid-1998, went back to the low-volatility phase in mid-2003, and
again switched back to the high-volatility phase in mid-2007.  The
first high-volatility phase observed in this period lasted five years,
within which we find not only the official March--November 2001
recession, but also the 2000 high in the DJI.  It is generally
believed that the DJI 2000 high is the result of the Dot-Com Bubble,
even though the March 2000 NASDAQ Crash did not even registered on the
DJI.  Very interestingly, apart from more or less isolated market
corrections, we find a series of market corrections which gets more
and more severe prior to the mid-1998 phase transition.  We realized
that these are precursor shocks similar in nature to those found by
Sornette \emph{et al} preceding market crashes
\cite{Sornette1996JPhysIFrance6p167, Sornette1997PhysicaA245p411,
Johansen1999EurPhysJB9p167, Johansen2000IntJTheorApplFin3p219}.  From
Figure \ref{fig:DJIsegments}, we see that the first precursor shock
appeared right after the July 1997 Asian Financial Crisis.  This
suggests, at least on face value, that the mid-1998 transition was
triggered by the Asian Financial Crisis.  Looking at the end of this
first high-volatility phase, we find a series of inverted shocks,
starting shortly after the DJI 2002 low.  Just like the precursor
shocks preceding the low-to-high transition, these low- to
moderate-volatily inverted shocks went on for about a year before the
US market made the high-to-low phase transition.  Though we do not yet
understand the nature of these shocks and inverted shocks, it is
likely that they are generic features in the dynamics of stock
markets.

The second high-volatility phase observed in the DJI time series is
none other than the present global financial crisis.  Depending on the
sources, the Subprime Crisis, which catalyzed the current global
financial crisis, is dated as early as July 2007.  On the surface,
there seems to be no connection between this gradual downturn, and
the Feb 2007 market crash known as the Chinese Correction.  However,
we find the Chinese Correction sitting in the middle of a year-long
precursor shock period starting in May 2006, marked by a less severe
market event that also had to do with corrections in the Chinese
markets.  Again, on face value, the US financial crisis appears to be
triggered by structural upheavals in a foreign market.  However, given
that US has substantial investment interests in China, it is not clear
from our observations what the true causes and effects might be.
Between September 2008 and April 2009, we have yet to detect any
inverted shocks, although it is likely the DJI has seen its lowest
point of this crisis in March 2009.  In the most optimistic scenario
that we start finding inverted shocks in April or May 2009, and
assuming the fundamental dynamics underlying these entities have not
changed from the previous crisis to the present crisis, we can expect
the US market to complete the high-to-low phase transition
(effectively an economic recovery) in mid-2010.

Finally, after learning so much from the DJI time series, it is
natural to ask if it is possible to avert an impending financial
crisis, if early detection based on precursor shocks is reliable.  To
answer such a question, we will need to understand the interplay
between factors that caused the precursor shocks.  At the very worst,
if we cannot understand the nature of these precursor shocks, they
would remain useful as early warning indicators of the financial
crisis.  Our hope then would be that intervention measures meted out
early may be able to soften the crisis, and perhaps even shorten it.
Equally important, if we can understand what we did in the previous
crisis that culminated in the inverted shocks, we might be able to
develop more systematic measures to aid recovery from the current
crisis.

\section{Conclusions}
\label{sect:conclusions}

We performed statistical segmentation of the DJI time series between
January 1997 and August 2008, using an optimized recursive
segmentation scheme derived from that introduced by Bernaola-Galv\'an
\emph{et al}.  We assumed normal as well as log-normal index movements
in each unknown statistically stationary segment of the time series,
and used the Jensen-Shannon divergence as the statistical distance
between segments.  Adopting the termination heuristic described in
Section \ref{sect:models}, we found 116 segments for the normal index
movement model, and 119 for the log-normal index movement model.
These two segmentations agree very well with each other, suggesting
that the segment boundaries discovered are statistically robust.  We
then performed agglomerative hierarchical clustering of the segments
using the complete-link algorithm, to find that the large number of
segments can be assigned to between five and seven clusters.  These
clusters are distinguished primarily by their variances, and represent
low-volatility, moderate-volatility, high-volatility, and
extremely-high-volatility macroeconomic phases.

Based on the temporal distribution of the clustered segments, we saw
that the US economy, as measured by the DJI, is found predominantly in
the low-volatility phase or the high-volatility phase.  The
low-volatility phase corresponds very roughly to the economic
expansion phase of the standard economic cycle.  In contrast, the
accepted economic contraction phase is completely nested within the much
longer high-volatility phase.  Both phases are interrupted frequently
by the week-long or month-long moderate-volatility market correction
phases.  Market crashes, which form a distinct macroeconomic phase
with extremely high volatility, occur with durations ranging from one
day to three weeks, and is almost exclusively found within the
high-volatility phase.  Within the period studied, we found the
high-volatility occuring only twice.  The first such interval was from
mid-1998 to mid-2003.  The second interval is the ongoing global
financial crisis which, according to our results, started in mid-2007.

From the temporal distribution of clustered segments, we found a
series of moderate-volatility precursor shocks preceding the mid-1998
low-to-high phase transition, and also a series of moderate-volatility
inverted shocks preceding the mid-2003 high-to-low phase transition,
which is associated with economic recovery that started with the DJI
2002 low.  There is also a series of precursor shocks preceding the
mid-2007 low-to-high phase transition that brought many financial
giants around the world to their knees.  The time scale for all
transitions identified from the DJI time series is about a year.  We
suspect inverted shocks would again appear roughly a year before the
end of the current financial crisis.  The implication of this finding
is that, if we do find inverted shocks trailing the the March 2009
low, and take these as the start of the economic recovery, the US
economy might find itself back in the low-volatility phase sometime in
the middle of 2010.  Should this optimistic scenario pan out, the
current high-volatility global financial crisis would have lasted
about three years, compared to five years for the previous
high-volatility phase.

From the DJI time series data alone, we see at face value that the
mid-1998 transition was triggered by the July 1997 Asian Financial
Crisis.  This assessment runs contrary to most accounts, because the
US market actually went on to scale new heights in 2000.  However,
because of the high volatility between 1998 and 2000, the upward trend
within this period must be interpreted very carefully.  In comparison,
the local trend between 2004 to 2007 is statistically much more
significant, because of the low volatility within this period.  While
the February 2007 market crash known as the Chinese Correction might
have played an important role, we see that there are earlier signs for
the start of global economic decline in mid-2007.  This is an unnamed
market event in May 2006, also related to correction within the
Chinese markets.  All in all, the story that unfolded from our
analysis of the DJI time series tells us how the global economies are
so coupled to each other, that structural transitions in one market
eventually propagates to most markets around the world.

Presently, we have initiated a comparative study of the Nikkei 225 and
the DJI over the same period (January 1997 to August 2008), to see
whether there are statistical signatures that point to causal links
between the US and Japanese markets.  At the same time, we are
replicating the analysis for the Dow Jones family of US economic
sector time series, to search for causal links between different
economic sectors.  We hope this more extensive analysis will tell us
which economic sectors follow which other economic sectors into
decline during a financial crisis.  We also hope to see which economic
sectors lead the economic recovery, and which economic sectors
are lifted up by others as the economy recovers.  Ultimately, a better
understanding of the causal relationships between economic sectors
will hint to more effective, and less costly stimulus measures.

\section*{Acknowledgements}

This research is supported by the Nanyang Technological University
startup grant SUG 19/07.  We have had helpful discussions with Low
Buen Sin, Charlie Charoenwong, Gerald Cheang Hock Lye, and Chris Kok
Jun Liang.

\end{document}